\documentclass[epj]{svjour}
\usepackage{graphics}
\begin{document}
\title{Behavior of the  topological susceptibility at finite T and $\mu$ and signs of restoration of chiral symmetries}
\titlerunning{Behavior of the  topological susceptibility at finite T and $\mu$...}
\author{M. C. Ruivo, 
				P. Costa
				\and C. A. de Sousa}
\institute{Centro de F\'{\i}sica Te\'{o}rica, Departamento de F\'{\i}sica, Universidade de Coimbra,
P-3004-516 Coimbra, Portugal}

%
\date{Received: date / Revised version: date}
%
\abstract{
We investigate the possible restoration of chiral and axial symmetries across the phase transition at finite temperature and chemical potential, by analyzing the behavior of several physics quantities, such as the quark condensates and the topological susceptibility,  the respective derivatives in order to chemical potential, and   the  masses of meson chiral partners. We discuss whether only chiral symmetry or both chiral and axial symmetries are restored and what is the role of the strange quark. The results are compared with recent lattice results.}

\PACS{
      {11.30.Rd}{ } \and
      {11.10.Wx}{ } \and
      {14.40.Aq}{ } 
			} 
%
\maketitle


Understanding the rich content of the QCD phase diagram   is a major challenge   nowadays. Phase transitions, associated to deconfinement, restoration of chiral and axial U$_A$(1) symmetries are expected to occur at high density and /or temperature. A question that has attracted a lot of attention is whether these phase transitions take place simultaneously and which observables could signal its occurrence. 

The U$_A$(1) symmetry is explicitly broken at the quantum level by the axial anomaly, that may be described at the semiclassical level by instantons, giving a mass to $\eta'$ in the chiral limit, which implies that, in the real world, this meson is not a remnant of a Goldstone boson. The U$_A$(1) anomaly causes flavor mixing, which has the effect of lifting the degeneracy between several mesons. So, the effective restoration of this symmetry should have relevant consequences on the meson masses as well as on the phenomenology of meson mixing angles. In particular, the $\eta'$  mass should decrease and this meson should degenerate with other Goldstone bosons. 
There are several reasons  to expect that the singlet axial symmetry might be restored. In fact, large instantons are supposed to be suppressed at high densities or temperatures,  and  interactions between instantons contribute to eliminate fluctuations of the topological charge, what implies that the effects of the anomaly could disappear \cite{Schaefer:2004}. The topological susceptibility, $\chi$, is related to the $\eta'$ mass through the Witten-Veneziano formula, and the behavior of $\chi$ and its slope are relevant to understand the possible restoration of the U$_A$(1) symmetry \cite{Meggiolaro:1992,Schafner:2000,Fukushima:2001,Costa:2004}.
 
The topological susceptibility is defined as:

\begin{equation}
 \chi=\int{\rm d}^4x\;\langle
 T \{Q(x)Q(0)\}\rangle,
\end{equation}

\noindent where $Q(x)$ is the topological charge density.  Several lattice calculations (see \cite{Alles:1997} and references therein) indicate a sharp decrease of this quantity with temperature at zero density. Model calculations also give indications in this direction. Chiral models include an anomaly term that breaks the $U_A(1)$ symmetry. In order to simulate the fate of the anomaly it is  usually assumed that the anomaly coefficient is a dropping function of temperature, whether the approach is phenomenological \cite{Alkofer:1989} or lattice inspired \cite{Schafner:2000,Fukushima:2001,Costa:2004}. 
Recent lattice calculations with two colors and eight flavors \cite{Alles:2006} show  that, at a fixed  $T$, and varying the chemical potential $\mu$,  a critical $\mu$ is found, where the quark condensate and the topological susceptibility drop and the Polyakov loop raises; its derivatives vary sharply.
The topological susceptibility and the quark condensate go to zero at much higher $\mu$.

Another subject that deserves attention is the role played by the strange quark regarding the restoration of chiral and axial symmetries. In fact, chiral symmetry could still be broken for strange quarks, while it is already restored for light quarks. Lattice calculations show that the critical temperature for light $\langle \bar q q\rangle$ is lower than for $\langle \bar s s\rangle$ \cite{Bernard:2005}, a result also found in chiral perturbation theory and model calculations \cite{Pelaez:2002,Toublan:2005}. However the critical temperatures could become closer  at high chemical potential \cite{Toublan:2005}. 

Here we report a study on  the behavior of the topological susceptibility at finite temperature, $T$, and baryonic chemical potential, $\mu_B= \mu_q=\mu_s$, and the possible restoration of chiral and axial symmetries, discussing, in particular, the role played by the strange quark in this concern. 

\begin{figure*}[t]
\begin{center}
\resizebox{0.32\textwidth}{!}{%
  \includegraphics{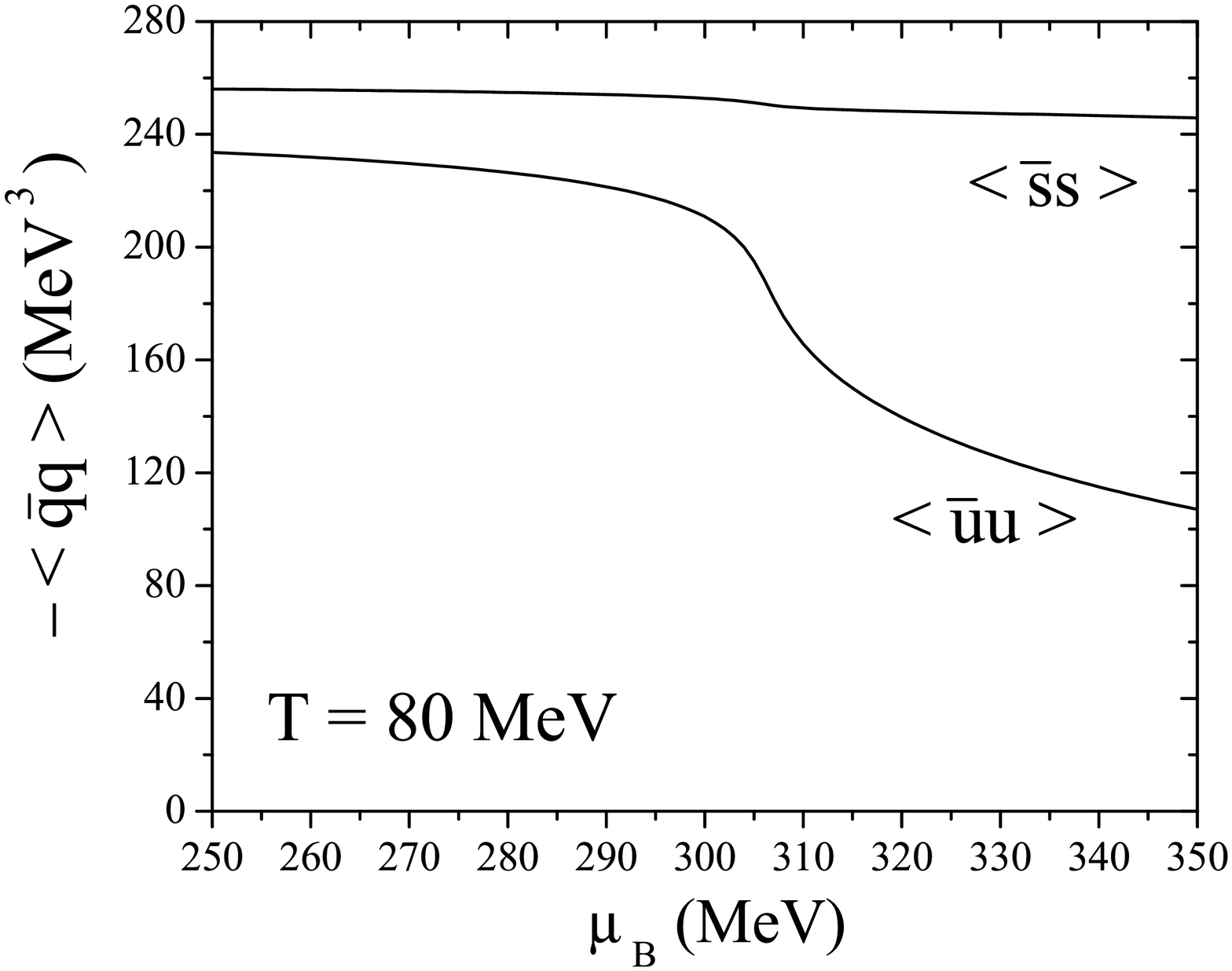}}
\resizebox{0.33\textwidth}{!}{%
  \includegraphics{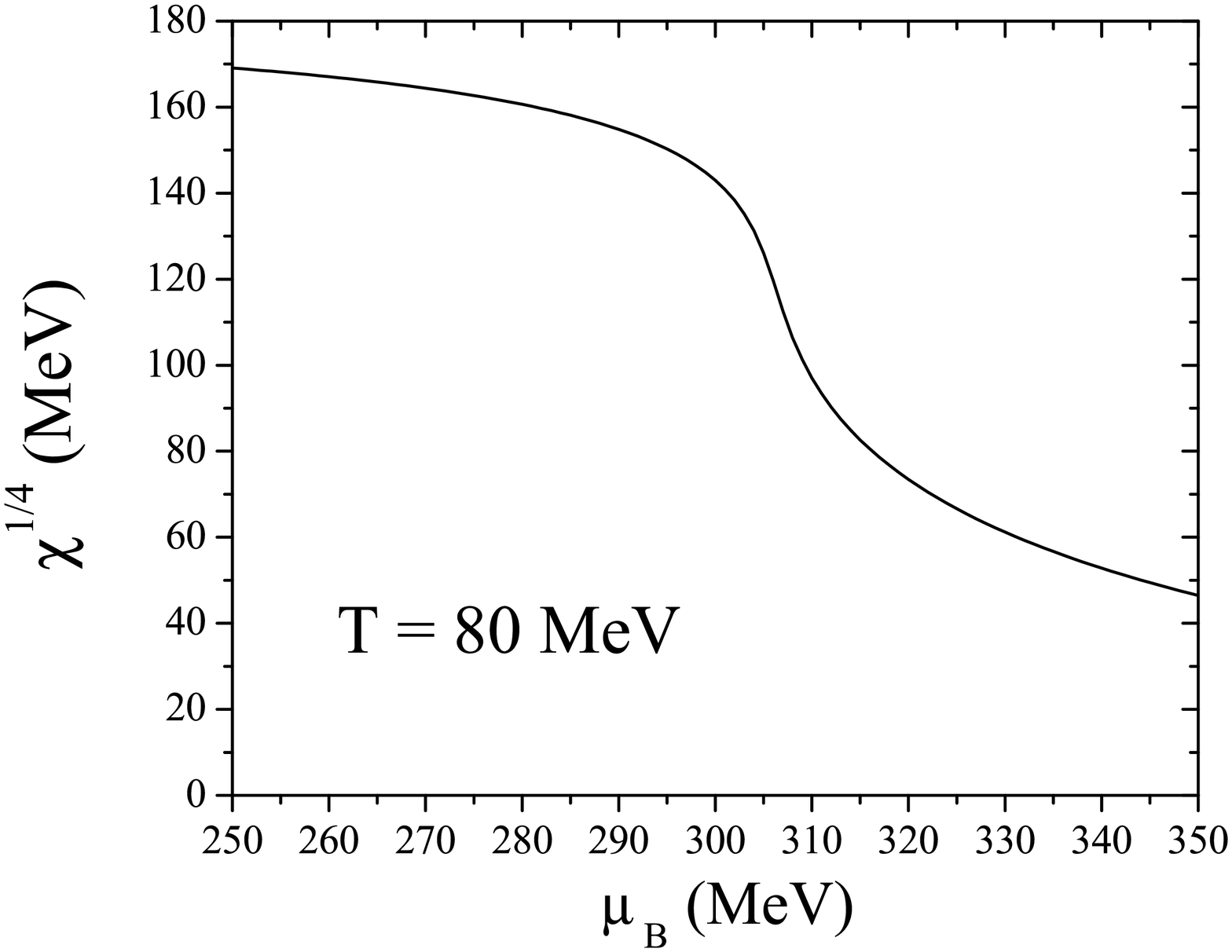}}
\resizebox{0.32\textwidth}{!}{%
	\includegraphics{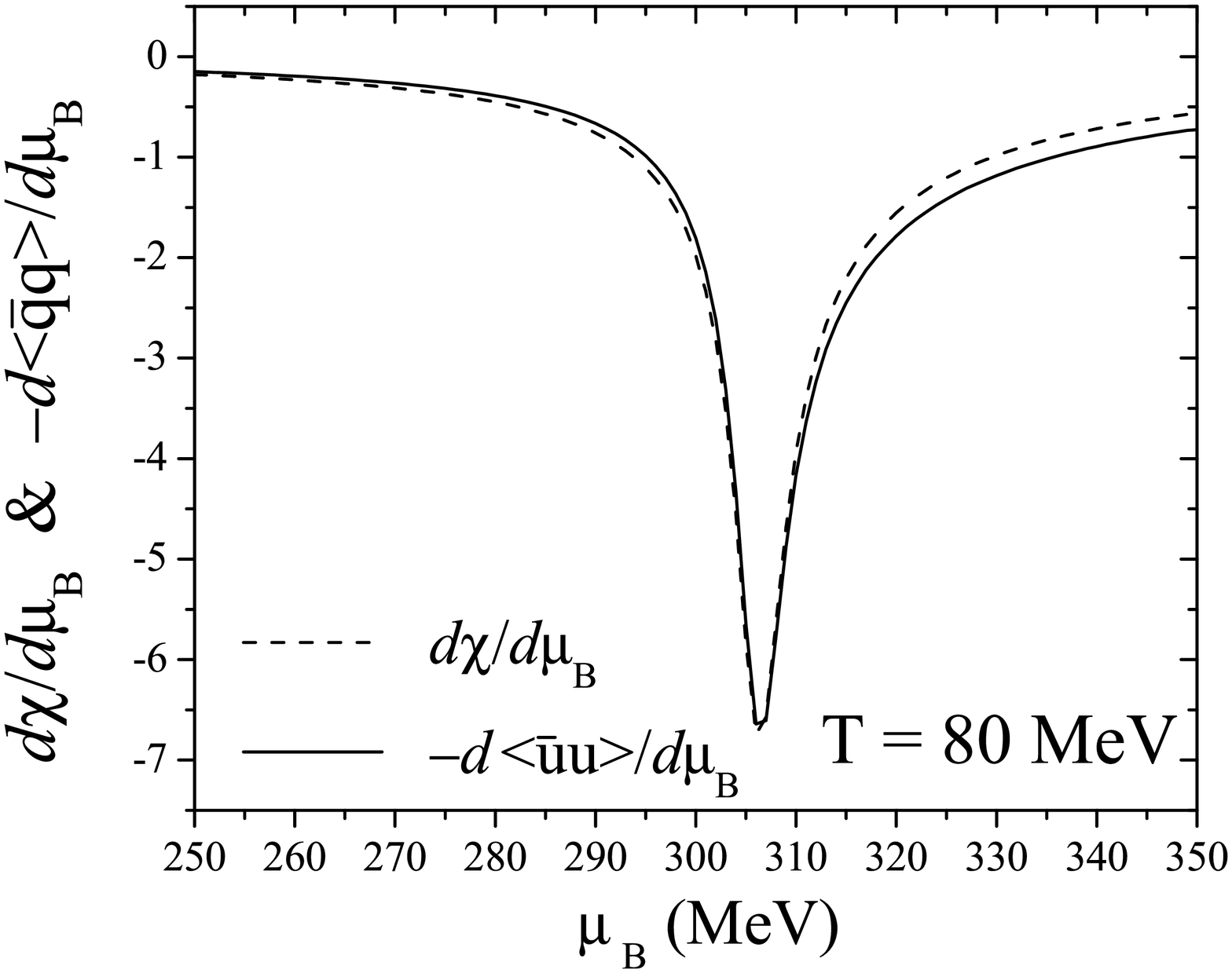}}
\end{center}  
\caption{Quark condensates (left panel), the topological susceptibility (midle panel) and the respective derivatives (right panel) (Case I).}
\label{fig:1}       
\end{figure*}

We perform our calculations in the framework of the  three--flavor  NJL model, including the determinantal 't Hooft interaction that breaks the U$_A$(1) symmetry, which has the  following Lagrangian: %
\begin{eqnarray} \label{lagr}
{\mathcal L} &=& \bar{q} \left( i \partial \cdot \gamma - \hat{m} \right) q
+ \frac{g_S}{2} \sum_{a=0}^{8}
\Bigl[ \left( \bar{q} \lambda^a q \right)^2+
\left( \bar{q} (i \gamma_5)\lambda^a q \right)^2
 \Bigr] \nonumber \\
&+& g_D \Bigl[ \mbox{det}\bigl[ \bar{q} (1+\gamma_5) q \bigr]
  +  \mbox{det}\bigl[ \bar{q} (1-\gamma_5) q \bigr]\Bigr] \, .
\end{eqnarray}
Here $q = (u,d,s)$ is the quark field with $N_f=3$  and $N_c=3$. $\hat{m}=\mbox{diag}(m_u,m_d,m_s)$ is the current 
quark mass matrix and $\lambda^a$ are the Gell--Mann matrices, 
a = $0,1,\ldots , 8$, ${ \lambda^0=\sqrt{\frac{2}{3}} \, {\bf I}}$. 
The model is fixed by the coupling constants $g_S, g_D$, the cutoff parameter $\Lambda$, which regularizes the divergent integrals, and the current quark masses $m_i$.
We use the parameter set: 
$m_u = m_d = 5.5$ MeV, $m_s = 140.7$ MeV, $g_S \Lambda^2 = 3.67$, $g_D \Lambda^5 = -12.36$ and $\Lambda = 602.3$ MeV (for details see \cite{Costa:2003}) .

In a  previous work  \cite{Costa:2004}, we studied the possible effective restoration of axial symmetry, at finite temperature and zero chemical potential, by modeling the anomaly coefficient  as a Fermi function from lattice results for $\chi$; a similar approach was extrapolated for quark matter at zero temperature and finite density.  For the sake of comparison with our present study, we summarize here the main conclusions for $\mu_B=0$ and $T\not=0$. It was found that at  $T\approx250$ MeV the  chiral partners  ($\pi^0,\sigma$) and ($a_0,\eta$) become degenerate, which is a manifestation of   the effective restoration of $SU(2)$ chiral symmetry; at  $T\approx350$ MeV,  $\chi \,\rightarrow\,0$, the pair ($a_0$,  $\sigma$)  becomes degenerate with ($\pi^0$,  $\eta$) and  the mixing angles go to the ideal values, indicating an effective restoration of axial symmetry. The strange quark condensate decreases slightly but, in the range of temperatures considered, chiral symmetry in the strange sector remains broken. The mesons $\eta'$ and  $f_0$ become purely strange and do not show a tendency to converge. Therefore there is no restoration of the full $U(3)\otimes U(3)$ symmetry.
  
We would like  to discuss questions such as whether the restoration of  chiral symmetry contributes to the restoration of the singlet axial symmetry U$_A$(1),  what is the role of the strange quark and  the combined effect of finite $T\,,\mu$ on the restoration of symmetries. For this purpose we will  study  the relevant observables  as functions of the chemical potential, $\mu_B=\mu_q=\mu_s$, at  a fixed temperature (below/above the critical end point (CEP))  in two cases: Case I: $m_q\not=m_s$; Case II: $m_q=m_s=0.5$ MeV. $g_D$ is kept constant in both cases. The main difference between results at a fixed $T$, below and above  the CEP  is that, in the first case,  the observables present a discontinuity at the critical $\mu_B$. 

Concerning Case I, we plot in Fig. 1 only the results above the  CEP ($\mu_B= 318.5$ MeV, $T=67.7$ MeV). It can be seen that there is a pronounced decrease of  $<\bar u u>$, but not of  $<\bar s s>$; $\frac{\partial <\bar u u>}{\partial \mu_B}\,\,,\frac{\partial \chi}{\partial \mu_B}$ vary sharply at the same critical $\mu_B$. We interpret this result as indicating a  phase transition associated to  partial restoration of chiral $SU(2)$. Using as a criterion for the effective restoration of symmetries the convergence of the respective chiral partners, we verified that the effective restoration of  chiral $SU(2)$ symmetry occurs at  a higher $\mu_B$, but there is no effective restoration of U$_A$(1) symmetry: the respective chiral partners do not converge and $\chi$ does not vanish. So, when $g_D$ is kept constant, the combined effect of finite $T$ and $\mu_B$ is not sufficient to lead to effective restoration of axial symmetry. 

In order to  compare our results with those from lattice calculations \cite{Alles:2006} and to discuss  the role of the strange quark, we will consider Case II (CEP : $\mu_B=219$ MeV, $T=87$ MeV). Since all the quark masses are equal, $a_0$ is degenerate with $\sigma$ and $\eta$ with $\pi^0$, even in the vacuum. 

In Fig. 2 we see that the quark condensate and the topological susceptibility have a behavior that compares well, at a qualitative level, with the results of \cite{Alles:2006}, as well as  the derivatives $\frac{\partial <\bar q q>}{\partial \mu_B}\,\,,\frac{\partial \chi}{\partial \mu_B}$, that  vary sharply at the same critical $\mu_B$  ($\mu_B\simeq 250 (200)$MeV for $T\simeq 50 (100)$MeV) (Fig. 3, left panel). This critical $\mu_B$ is certainly the onset of a new phase with partial restoration of  chiral symmetry. The information from the mesonic behavior (Fig. 3, right panel) shows that,differently from Case I, the effective and partial restoration of chiral symmetry occur at  about the the same critical chemical potential (convergence of the chiral partners ($\pi^0,\sigma$)). However, the  effective restoration of U$_A$(1) symmetry  takes place at a higher chemical potential ($\mu_B\simeq 350$ MeV for $T\simeq 100$ MeV): the pair ($f_0,\eta'$) degenerates with ($\pi^0,\sigma$) and $\chi$ vanishes.  
For higher temperatures, the phase transition occurs earlier. 

\begin{figure*}[t]
\begin{center}
\resizebox{0.4\textwidth}{!}{%
  \includegraphics{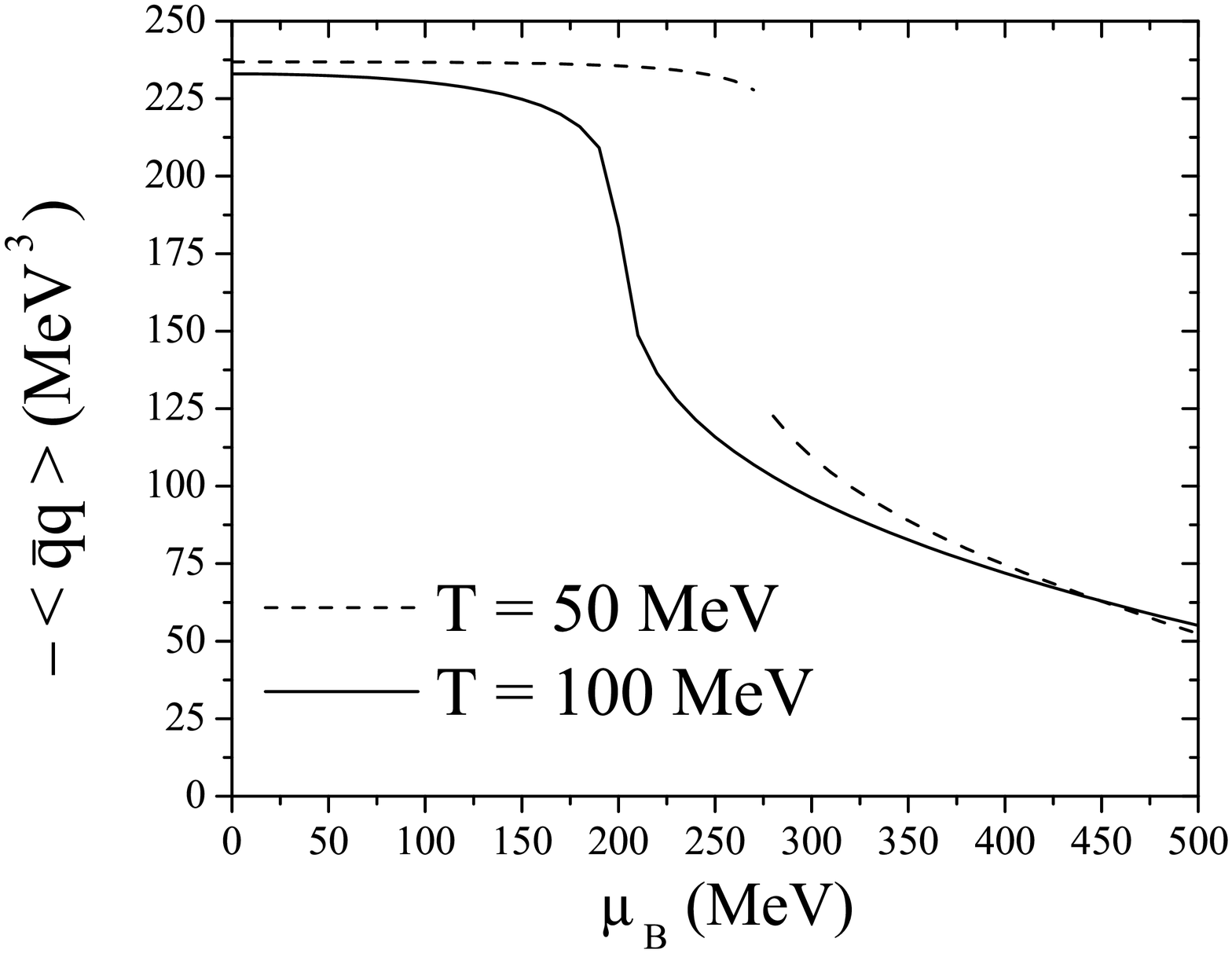}}
\resizebox{0.43\textwidth}{!}{%
  \includegraphics{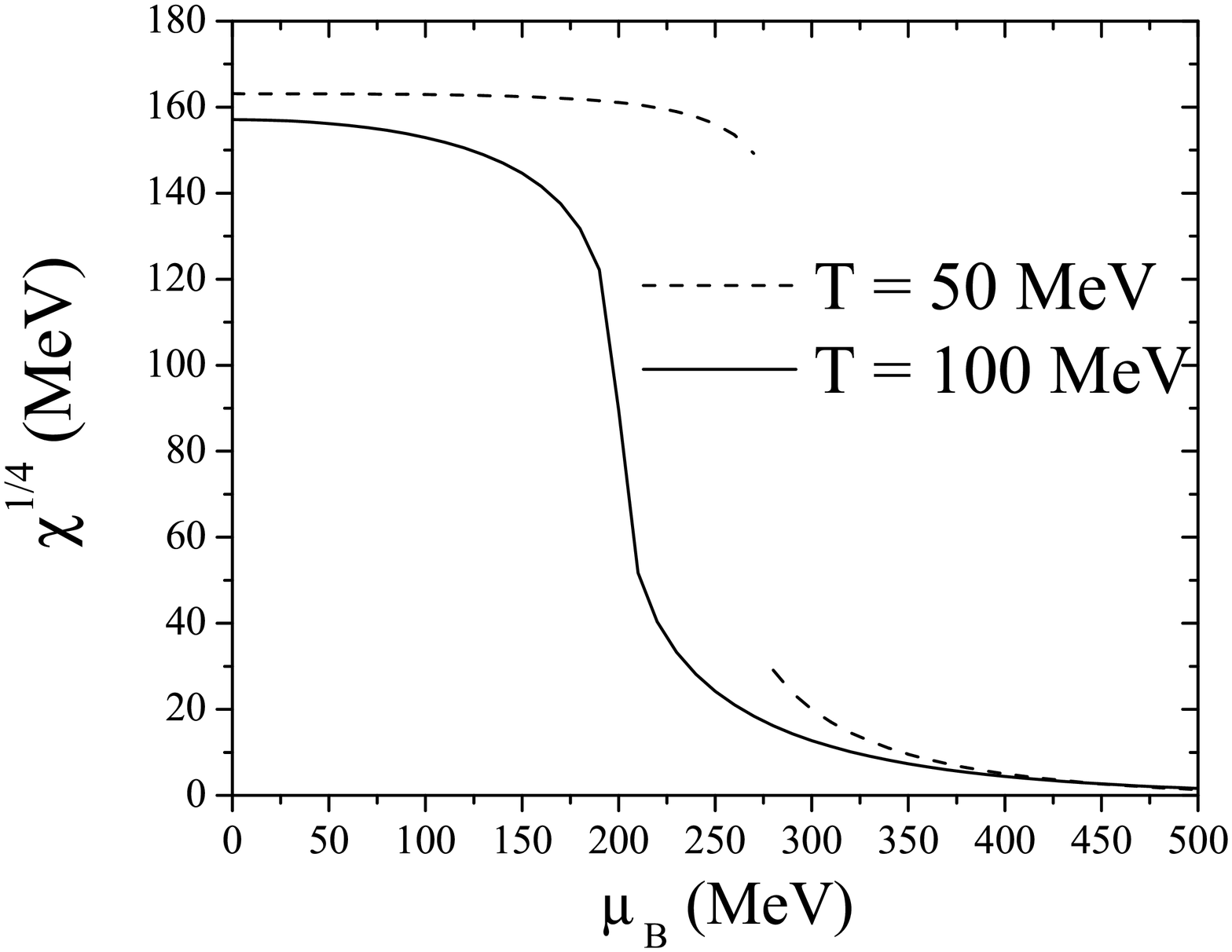}}
\end{center} 
\caption{Quark condensates (left panel) and topological susceptibility (right panel), above the CEP (Case II).}
\label{fig:3}       
\end{figure*}

\begin{figure*}[t]
\begin{center}
\resizebox{0.4\textwidth}{!}{%
  \includegraphics{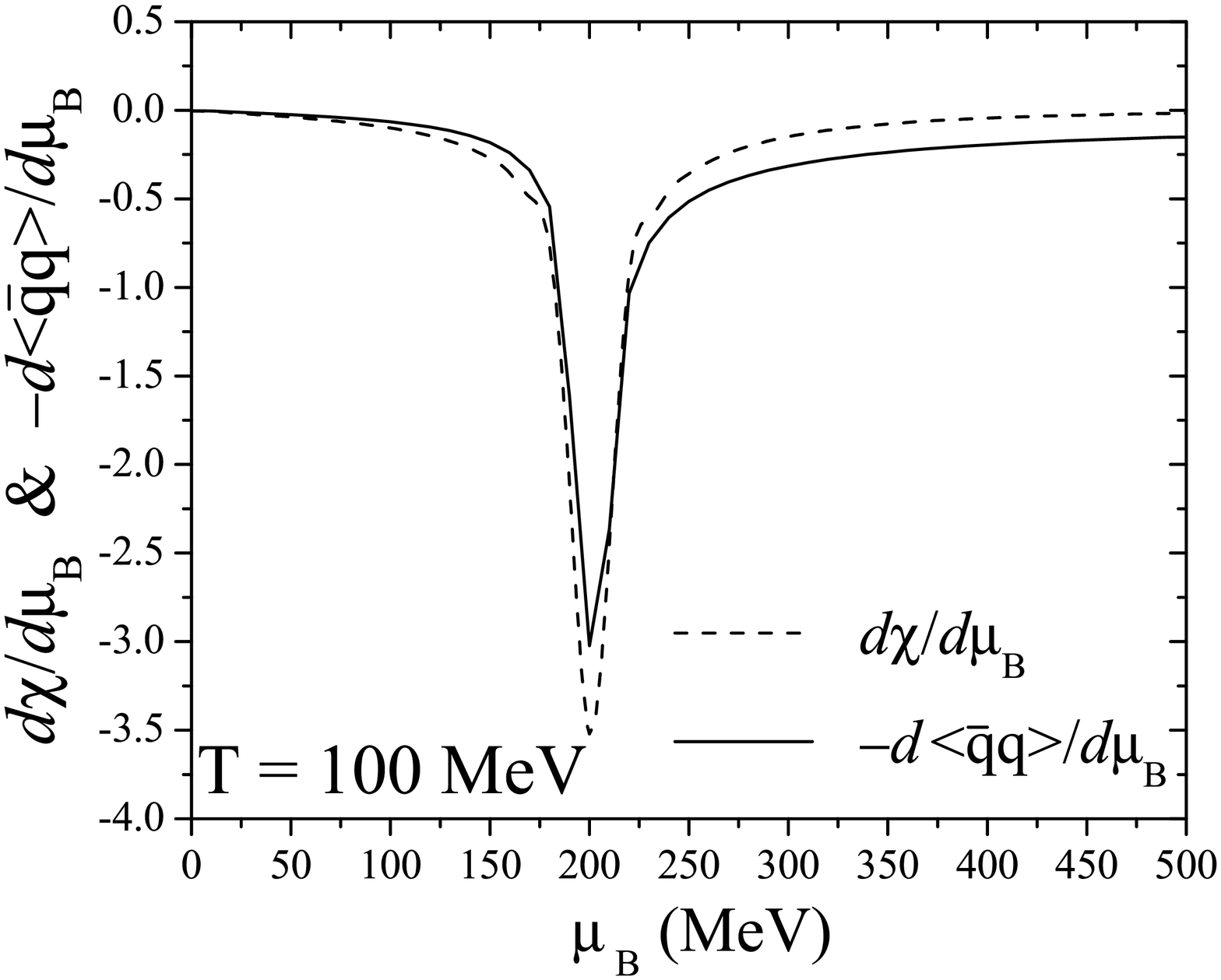}}
\resizebox{0.41\textwidth}{!}{%
  \includegraphics{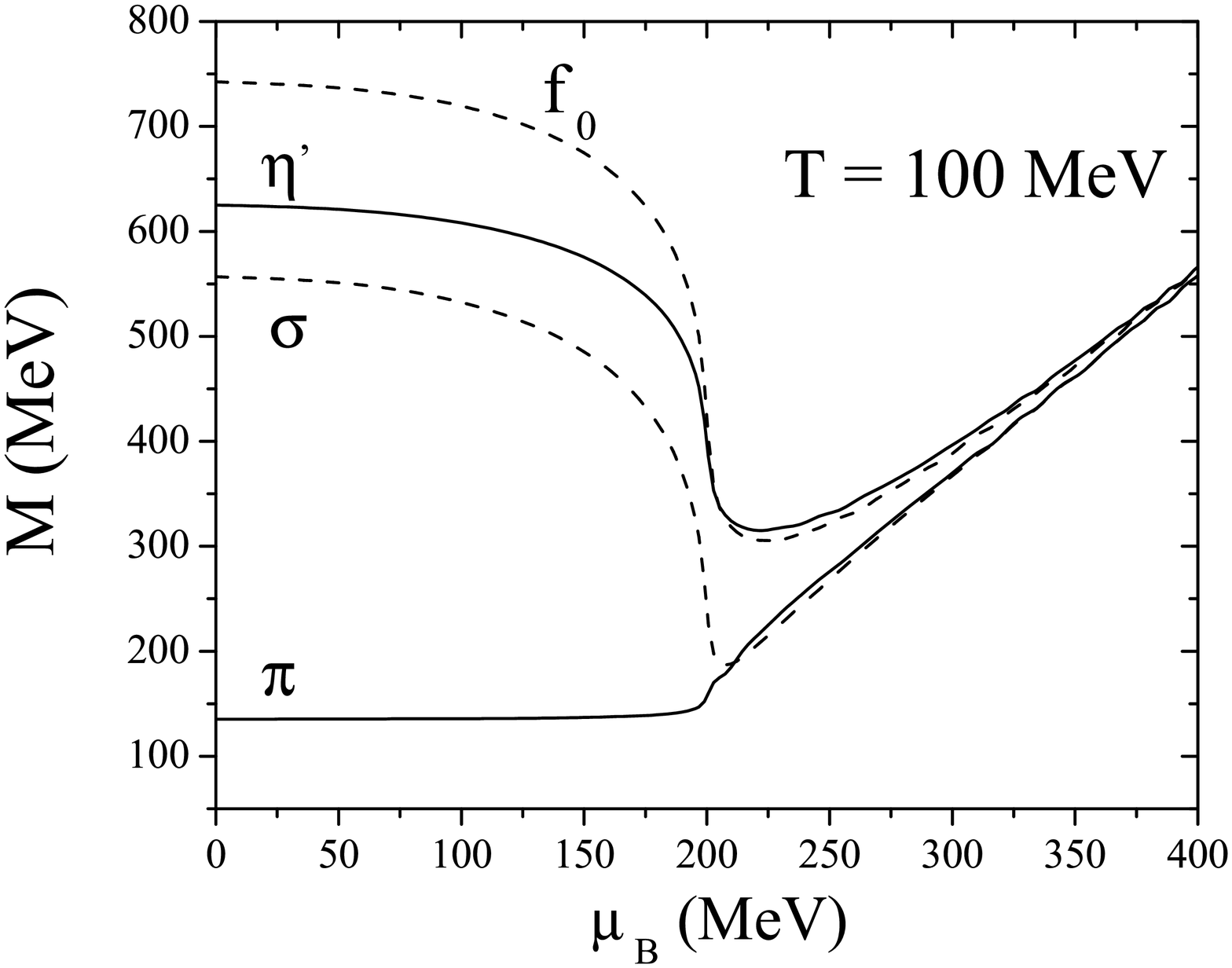}}
\end{center}  
\caption{Derivatives of $<\bar q q>$ and $\chi$ (left panel) and meson masses (right panel), above the CEP (Case II).}
\label{fig:4}       
\end{figure*}

In conclusion,  we have  studied the behavior of the topological susceptibility  across the phase transition  and discussed its relation with the restoration of chiral symmetries. We studied the observables  as functions of the chemical potential, $\mu_B=\mu_q=\mu_s$, at  a fixed temperature (below/above the critical end point (CEP)) for two cases. 
In Case I, the only way of obtaining effective restoration of axial symmetry is by  assuming that $g_D$ is a dropping function of $T (\mu)$. In Case II, we have effective restoration of axial symmetry with  $g_D=$Cte : $\chi$ vanishes and all the mesons converge; The difference between the results for the two cases is due to the behavior of the strange quark condensate.	

\vspace{0.5cm}
Work supported by grant SFRH/BPD/23252/2005 from FCT (P. Costa).




\end{document}